\documentclass[twoside,12pt]{article}
\usepackage{epsfig}

\newcommand{\be}{\begin{equation}}
\newcommand{\ee}{\end{equation}}
\newcommand{\bea}{\begin{eqnarray}}
\newcommand{\eea}{\end{eqnarray}}

\begin{document}

\title{Strangeness dynamics in relativistic nucleus-nucleus
collisions\thanks{Supported by DFG, GSI, BMBF and DESY} }
\author{E. L.\ Bratkovskaya$^{1}$,
M.\ Bleicher$^1$, W.\ Cassing$^2$, M.\ van Leeuwen$^3$, \\
M.\ Reiter$^1$, S.\ Soff$^1$, H.\ St\"ocker$^1$, H. Weber$^1$ \\ \\
$^1$Institut f\"{u}r Theoretische Physik,
   Universit\"{a}t Frankfurt,  Germany\\
$^2$Institut f\"{u}r Theoretische Physik,
   Universit\"{a}t Giessen,  Germany\\
$^3$NIKHEF, Amsterdam, Netherlands}
\date{ }
\maketitle

\begin{abstract}
We investigate  hadron production as
well as transverse hadron spectra in nucleus-nucleus collisions from 2
$A\cdot$GeV to 21.3 $A\cdot$TeV within two independent transport
approaches (UrQMD and HSD) that are based on quark, diquark, string and
hadronic degrees of freedom.  The comparison to experimental data
demonstrates that both approaches agree quite well with each other and
with the experimental data on hadron production.
The enhancement of pion production in central Au+Au (Pb+Pb)
collisions relative to scaled $pp$ collisions (the 'kink')
is well described by both approaches without involving any phase transition.
However, the maximum in the $K^+/\pi^+$ ratio at 20 to 30 A$\cdot$GeV
(the 'horn') is missed by $\sim$ 40\%.
A comparison to the transverse mass spectra from $pp$ and C+C
(or Si+Si) reactions shows the reliability of the transport models for
light systems. For central Au+Au (Pb+Pb) collisions at bombarding
energies above $\sim$ 5 A$\cdot$GeV, however, the measured $K^{\pm}$
$m_{T}$-spectra have a larger inverse slope parameter than expected
from the calculations. The approximately constant slope of $K^\pm$
spectra at SPS (the 'step') is not reproduced either.
Thus the pressure generated by hadronic interactions in the transport
models above $\sim$ 5 A$\cdot$GeV is lower than observed in the
experimental data. This finding suggests that the additional pressure -
as expected from lattice QCD calculations at finite quark chemical
potential and temperature - might be generated by strong interactions in the
early pre-hadronic/partonic phase of central Au+Au (Pb+Pb) collisions.
\end{abstract}

%\eject
%\tableofcontents
\section{Introduction}

The phase transition from partonic degrees of freedom (quarks and
gluons) to interacting hadrons is a central topic of modern high-energy
physics. In order to understand the dynamics and relevant scales of
this transition laboratory experiments under controlled conditions are
presently performed with ultra-relativistic nucleus-nucleus collisions.
Hadronic spectra and relative hadron abundancies from these experiments
reflect  important aspects of the dynamics in the hot and dense zone
formed in the early phase of the reaction.  Furthermore, as has been
proposed early by Rafelski and M\"uller \cite{Rafelski} the strangeness
degree of freedom might play an important role in distinguishing
hadronic and partonic dynamics.

In fact, estimates based on the Bjorken formula \cite{bjorken} for the
energy density achieved in central Au+Au collisions suggest that the
critical energy density for the formation of a quark-gluon plasma (QGP)
is by far exceeded during a few fm/c in the initial phase of the
collision at Relativistic Heavy Ion Collider (RHIC) energies
\cite{QM01}, but sufficient energy densities ($\sim$ 0.7-1 GeV/fm$^3$
\cite{Karsch}) might already be achieved at Alternating Gradient
Synchrotron (AGS) energies of $\sim$ 10 $A\cdot$GeV \cite{HORST,exita}.
More recently, lattice QCD calculations at finite temperature and quark
chemical potential $\mu_q$ \cite{Fodor} show a rapid increase of the
thermodynamic pressure $P$ with temperature above the critical
temperature $T_c$ for a phase transition to the QGP. The crucial
question is, however, at what bombarding energies the conditions  for
the phase transition (or cross-over)  might be fulfilled.

Presently,  transverse mass (or momentum) spectra of hadrons are in the
center of interest. It is experimentally observed that the
transverse mass  spectra of kaons at AGS and SPS energies show a
substantial {\it flattening} or {\it hardening}  in central Au+Au
collisions relative to $pp$ interactions (cf.
Refs.~\cite{NA49_T,Goren}).  In order to quantify this effect, the
spectra are often parametrised as:
\begin{eqnarray}
\label{slope}
\frac{1}{m_T} \frac{dN}{dm_T} \sim \exp(-\frac{m_T}{T})
\end{eqnarray}
where $m_T=\sqrt{m^2+p_T^2}$ is the transverse mass and $T$ is the
inverse slope parameter. This hardening of the spectra is commonly
attributed to strong collective flow, which is absent in the
$pp$ or $pA$ data.

The authors of Refs. \cite{SMES,Marek} have proposed to interpret the
approximately constant $K^\pm$ slopes above $\sim 30$ A$\cdot$GeV -- the
'step' -- as an indication for a phase transition
following an early suggestion by Van Hove \cite{Hove}.  This
interpretation is also based on a rather sharp maximum in the
$K^+/\pi^+$ ratio at $\sim$ 20 to 30 A$\cdot$GeV in central
Au+Au (Pb+Pb) collisions (the 'horn' \cite{SMES,Marek}).  However,
it is presently not clear, if the statistical model assumptions invoked
in Refs. \cite{SMES,Marek} hold to be reliable.

We will demonstrate in this contribution that the pressure needed to
generate a large collective flow -- to explain the hard slopes of the
$K^\pm$ spectra as well as the 'horn' in the $K^+/\pi^+$ ratio -- is not
produced in the present models by the interactions of hadrons in the
expansion phase of the hadronic fireball.  In our studies we use two
independent transport models that employ hadronic and string degrees of
freedom, i.e., UrQMD (v. 1.3) \cite{UrQMD1,UrQMD2} and HSD
\cite{Geiss,Cass99}. They take into account the formation and multiple
rescattering of hadrons and thus dynamically describe the generation of
pressure in the hadronic expansion phase. This involves also
interactions of 'leading' pre-hadrons that contain a valence quark
(antiquark) from a primary 'hard' collision (cf. Refs.  \cite{Geiss,Weber02}).

The UrQMD transport approach \cite{UrQMD1,UrQMD2} includes all baryonic
resonances up to  masses of 2 GeV as well as mesonic resonances up to
1.9 GeV as tabulated by the Particle Data Group \cite{PDG}. For
hadronic continuum excitations a string model is used with hadron
formation times in the order of 1-2~fm/c depending on the momentum and
energy of the created hadron.
In the HSD approach nucleons, $\Delta$'s, N$^*$(1440), N$^*$(1535),
$\Lambda$, $\Sigma$ and $\Sigma^*$ hyperons, $\Xi$'s, $\Xi^*$'s and
$\Omega$'s  as well as their antiparticles are included on the baryonic
side whereas the $0^-$ and $1^-$ octet states are included in the
mesonic sector. High energy inelastic hadron-hadron collisions in HSD are
described by the FRITIOF string model \cite{LUND} whereas low energy
hadron-hadron collisions are modeled based on experimental cross
sections. Both transport approaches reproduce the nucleon-nucleon,
meson-nucleon and meson-meson cross section data in a wide kinematic
range.  We point out, that no explicit parton-parton scattering
processes (beyond the interactions of 'leading' quarks/diquarks) are
included in the studies below contrary to the multi-phase transport
model (AMPT) \cite{Ko_AMPT}, which is currently employed from upper SPS
to RHIC energies.

\section{Hadron excitation functions and ratios}

\subsection{$p p$ versus central $A A$ reactions -- the 'kink'}

\begin{figure}[t]
\begin{center}
\begin{minipage}[l]{12 cm}
\epsfig{file=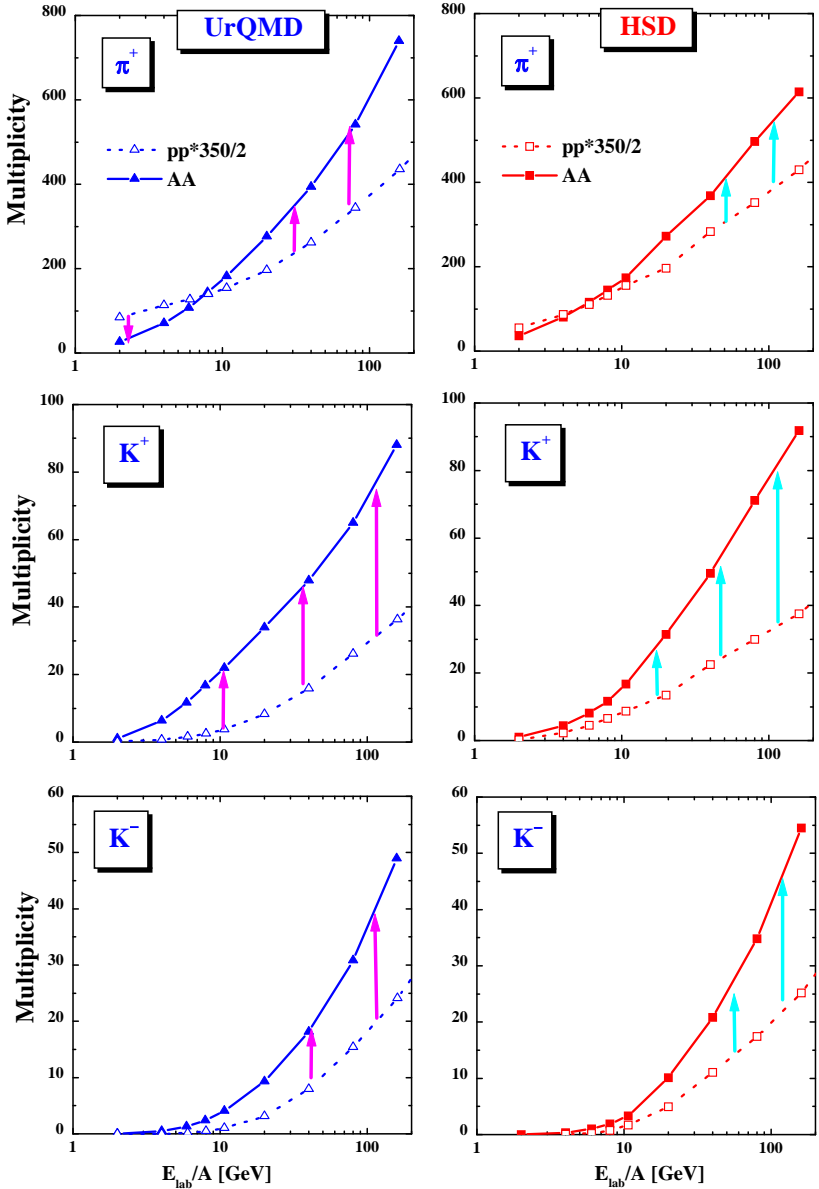,scale=0.6}
\end{minipage}
\phantom{a}\begin{minipage}[l]{5.5cm}
\caption{Total multiplicities of $\pi^+, K^+$ and
$K^-$ (i.e., $4\pi$ yields) from central Au+Au (at AGS) or Pb+Pb
(at SPS) collisions in comparison to the total
multiplicities from $pp$ collisions (scaled by a factor 350/2)
versus kinetic energy $E_{\rm lab}$. The solid lines with full
triangles and squares show the UrQMD (l.h.s.) and HSD results
(r.h.s.) for $AA$ collisions, respectively. The dotted lines with
open triangles and squares correspond to the $pp$ multiplicities
calculated within UrQMD (l.h.s.) and HSD (r.h.s.).
The figure is taken from Ref. \protect\cite{Weber02}.}
\label{multppaa}
\end{minipage}
\end{center}
\end{figure}

In order to explore the main physics from central $A A$ reactions
it is instructive to have a look at the various particle
multiplicities relative to scaled $pp$ collisions as a function of
bombarding energy. For this aim we show in Fig. \ref{multppaa}  the
total multiplicities of $\pi^+, K^+$ and $K^-$ (i.e., the $4\pi$
yields) from central Au+Au (at AGS) or Pb+Pb (at SPS)
collisions (from UrQMD and HSD) in comparison to the scaled
total multiplicities from $pp$
collisions versus the kinetic energy per particle $E_{\rm lab}$.

The general trend from both transport approaches is quite similar:
we observe a slight absorption of pions at lower bombarding energy
and a relative enhancement of pion production by rescattering
in heavy-ion collisions above $\sim$10 A$\cdot$GeV. Kaons and antikaons from
$AA$ collisions are always enhanced in central reactions relative
to scaled $pp$ multiplicities, which is a consequence of strong final
state interactions.  Thus, the 'kink' in the pion ratio as well as the
$K^\pm$ enhancement might result from conventional hadronic final state
interactions.

\subsection{Particle yields in central collisions of heavy nuclei}

\begin{figure}[th]
\begin{center}
\begin{minipage}[l]{12cm}
\epsfig{file=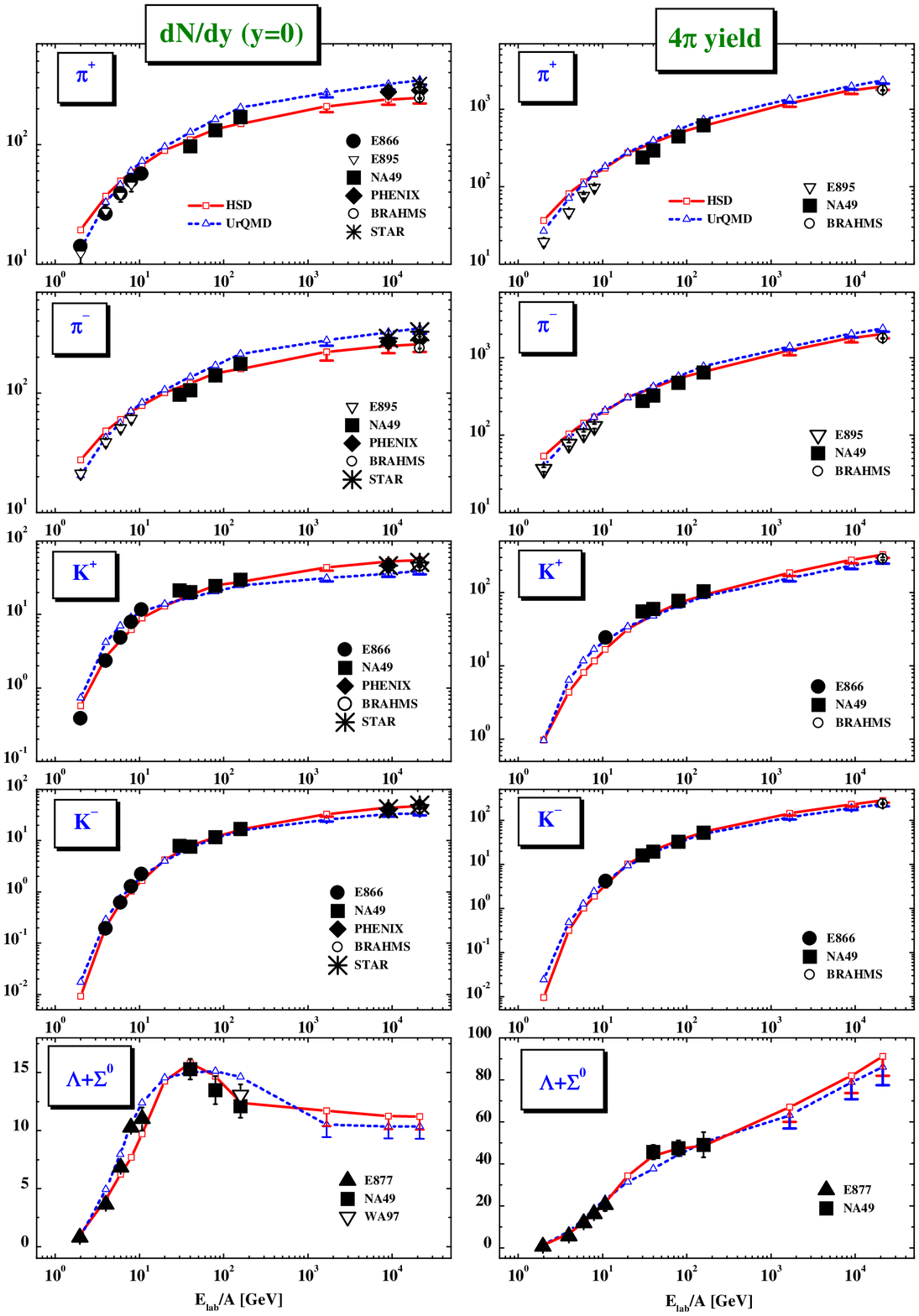,scale=0.63}
\end{minipage}
\hfill\phantom{a}\begin{minipage}[l]{5.5cm}
\caption{The excitation function of $\pi^+, \pi^-, K^+, K^-$ and
$\Lambda+\Sigma^0$ yields from 5\% central (AGS energies, SPS at 160
A$\cdot$GeV and at RHIC energies), 7\% central (20, 30, 40 and 80
A$\cdot$GeV), 10\% central for $\Lambda+\Sigma^0$ at 160 A$\cdot$GeV
Au+Au (AGS and RHIC) or Pb+Pb (SPS) collisions in comparison to the
experimental data from Refs.  \protect\cite{E866E917,E895,E891Lam}
(AGS), \protect\cite{NA49_new,NA49_Lam,Antiori} (SPS)
and \protect\cite{BRAHMS,PHENIX,STAR} (RHIC) for midrapidity
(left column) and rapidity integrated yields (right column).  The solid
lines with open squares show the results from HSD whereas the dashed
lines with open triangles indicate the UrQMD calculations. The lower
theoretical errorbars at RHIC energies correspond to the yields for
10\% central events. The fi\-gure is taken from Ref.
\protect\cite{Bratnew}.}
\label{Fig_yield}
\end{minipage}
\end{center}
\end{figure}
Fig. \ref{Fig_yield} shows the excitation function of $\pi^+, \pi^-,
K^+, K^-$ and $\Lambda+\Sigma^0$ yields (midrapidity (l.h.s.) and rapidity
integrated (r.h.s)) from central Au+Au (Pb+Pb) collisions in
comparison to the experimental data
\footnote{Note that all data from the NA49 Collaboration at 30 A$\cdot$GeV
have to be considered as 'preliminary'}.
As can be seen from Fig. \ref{Fig_yield} the differences between the
independent transport models are less than 20\%.  The maximum
deviations between the models and the experimental data are less than
$\sim 30$\%. In addition, a systematic analysis of the results from both
models and experimental data for central nucleus-nucleus collisions
from 2 to 160 $A\cdot$GeV in Ref. \cite{Weber02} has shown that also
the 'longitudinal' rapidity distributions of protons, pions, kaons,
antikaons and hyperons are quite similar in both models and in
reasonable agreement with available data.  The exception are the pion
rapidity spectra at the highest AGS energy and lower SPS energies,
which are overestimated by both models \cite{Weber02}.  For a more
detailed comparison of HSD and UrQMD calculations with experimental
data at RHIC energies we refer the reader to Refs.
\cite{Brat03,ERICE2,Soff03}.

\subsection{Particle ratios -- the 'horn'}

\begin{figure}[t]
\begin{center}
\begin{minipage}[l]{12.5cm}
\epsfig{file=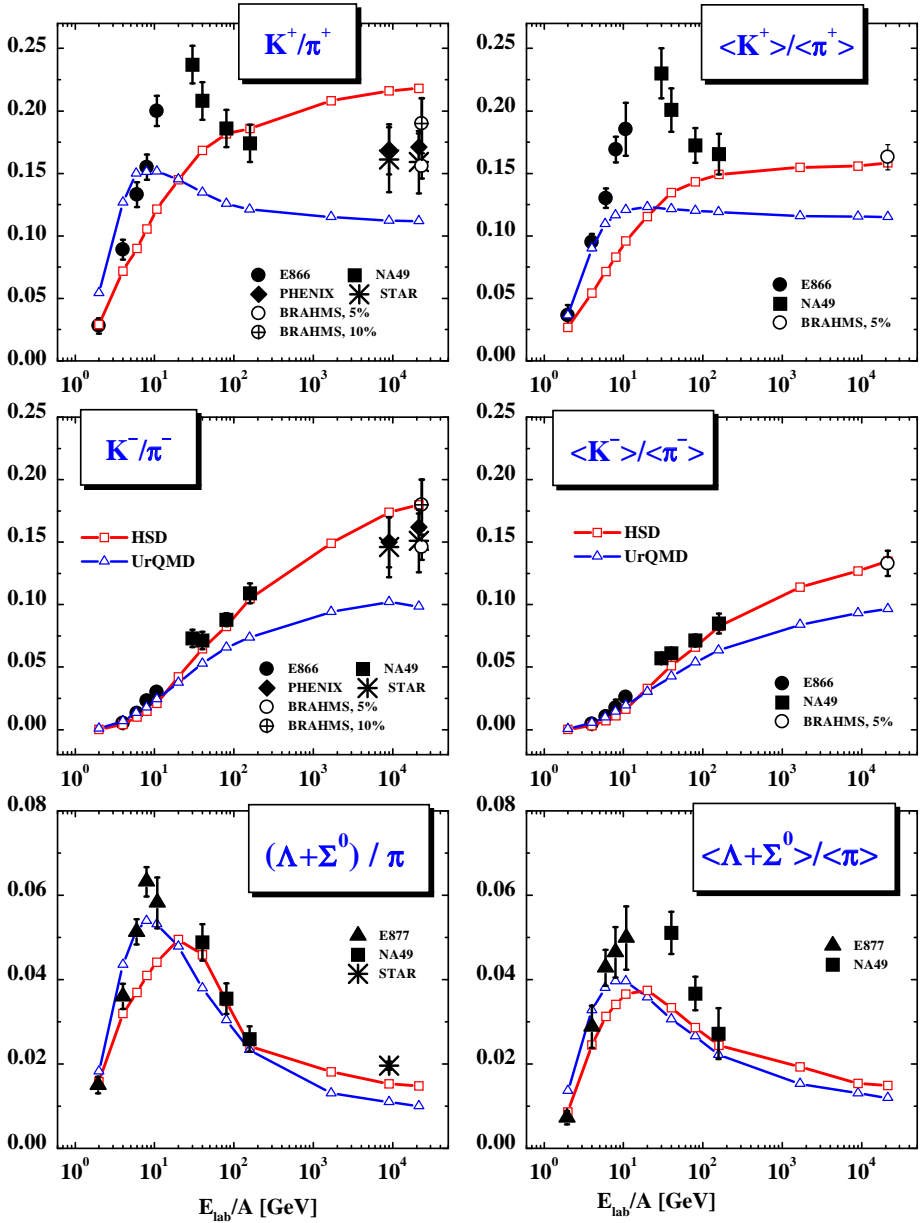,scale=0.65}
\end{minipage}
\phantom{a}\begin{minipage}[l]{5.cm}
\caption{The excitation function of $K^+/\pi^+, K^-/\pi^-$ and
$(\Lambda+\Sigma^0)/\pi$ ratios from 5\% central (AGS energies, SPS at 160
A$\cdot$GeV and at RHIC energies), 7\% central (20, 30, 40 and 80
A$\cdot$GeV), 10\% central for $\Lambda+\Sigma^0$ at 160 A$\cdot$GeV
Au+Au (AGS and RHIC) or Pb+Pb (SPS) collisions in comparison to the
experimental data from Refs. \protect\cite{E866E917,E891Lam}
(AGS), \protect\cite{NA49_new,NA49_Lam,Antiori} (SPS) and
\protect\cite{BRAHMS,PHENIX,STAR} (RHIC) for midrapidity (left column)
and rapidity integrated yields (right column).  The solid lines with
open squares show the results from HSD whereas the dashed lines with
open triangles indicate the UrQMD calculations.
The figure is taken from Ref. \protect\cite{Bratnew}.}
\label{Fig_rat}
\end{minipage}
\end{center}
\end{figure}
In Fig. \ref{Fig_rat} we present the excitation function of the
particle ratios $K^+/\pi^+, K^-/\pi^-$ and $(\Lambda+\Sigma^0)/\pi$
from central Au+Au (Pb+Pb) collisions in comparison to
experimental data.
The deviations between the transport models and the data are most
pronounced for the midrapidity ratios (left column) since the
ratios are very sensitive to actual rapidity spectra. The
$K^+/\pi^+$ ratio in UrQMD shows a maximum at $\sim$ 8 A$\cdot$GeV
and then drops to a constant ratio of 0.11 at top SPS and RHIC
energies. In the case of HSD a continuously rising ratio with
bombarding energy is found for the midrapidity ratios which partly
is due to a dip in the pion pseudo-rapidity distribution at
RHIC energies (cf. Fig. 1 in Ref. \cite{Brat03}). The 4$\pi$ ratio
in HSD is roughly constant  from top SPS to RHIC energies,
however, larger than the ratio from UrQMD due to the lower amount
of pion production\footnote{The lower amount of pions in HSD is
essentially due to an energy-density cut of 1 GeV/fm$^3$, which
does not allow to form hadrons above this critical energy
density \cite{Weber02}.}
and a slightly higher $K^+$ yield (cf. Fig. \ref{Fig_yield}).
Nevertheless, the experimental maximum in the $K^+/\pi^+$ ratio is
missed,  which we address dominantly to the excess of pions
in the transport codes rather than to missing strangeness production.
Qualitatively, the same arguments - due to strangeness conservation -
also hold for the $(\Lambda +\Sigma^0)/\pi$ ratio, where the pronounced
experimental maxima are underestimated due to the excess of pions in
the transport models at top AGS energies (for HSD) and above $\sim$ 5
A$\cdot$GeV (for UrQMD). Since the $K^-$ yields are well reproduced by
both approaches (cf. Fig. \ref{Fig_yield}) the deviations in
the $K^-/\pi^-$ ratios at SPS and RHIC energies in UrQMD can be traced back
to the excess of pions (see discussion above).

We stress that the maximum in the $(\Lambda +\Sigma^0)/\pi$ ratio is
essentially due to a change from baryon to meson dominated dynamics
with increasing bombarding energy. Similar arguments hold for the
experimentally observed maxima in the ratio $\Xi/\pi$  (cf. Ref.
\cite{Xipi}). However, the 'horn' in the $K^+/\pi^+$ ratio at
$\sim$30 A$\cdot$GeV is not described by both transport models.

\section{Transverse mass spectra -- the 'step'}

We now focus on transverse mass spectra of pions and kaons/antikaons
from central Au+Au (Pb+Pb) collisions from 2 $A\cdot$GeV to 21.3
$A\cdot$TeV  and compare to recent data (cf. Ref. \cite{Brat03PRL}).
Without explicit representation we mention that the agreement between
the transport calculations and the data for $pp$ and for central C+C and
Si+Si is quite satisfactory \cite{Brat03PRL}; no obvious traces of
'new' physics are visible.  The situation, however, changes for central
Au+Au (or Pb+Pb) collisions.  Whereas at the lowest energy of 4
$A\cdot$GeV the agreement between the transport approaches and the data
is still acceptable, severe deviations are visible in the $K^\pm$
spectra at SPS energies of 30 and 160 $A\cdot$GeV \cite{Brat03PRL}.  We
note that the $\pi^{\pm}$ spectra are reasonably described at all
energies while the inverse slope $T$ of the $K^\pm$ transverse mass
spectra in Eq. (\ref{slope}) is underestimated severely by about the
same amount in both transport approaches (within statistics).  The
increase of the inverse $K^\pm$ slopes in heavy-ion collisions with
respect to $pp$ collisions, which is generated by rescatterings of
produced hadrons in the transport models, is small because the
elastic meson-baryon scattering is strongly forward peaked and
therefore gives little additional transverse momentum at midrapidity.

The question remains whether the underestimation of the $K^\pm$ slopes
in the transverse mass spectra \cite{Brat03PRL} might be due to
conventional hadronic medium effects.  In fact, the $m_T$ slopes of
kaons and antikaons at SIS energies (1.5 to 2 $A\cdot$GeV) were found
to differ significantly \cite{KaoS}. As argued in  \cite{Cass99}
the different slopes could be traced back to repulsive kaon-nucleon
potentials, which lead to a hardening of the $K^+$ spectra,
and attractive antikaon-nucleon potentials, which lead to a softening
of the $K^-$ spectra. However, the effect of such potentials was
calculated within HSD and  found to be of minor importance at AGS and
SPS energies \cite{Cass99} since the meson densities are comparable to
or even larger than the baryon densities at AGS energies and above.
Additional self energy contributions stem from $K^\pm$ interactions
with mesons; however, $s$-wave kaon-pion interactions are weak due to
chiral symmetry arguments and $p$-wave interactions such as $\pi+K
\leftrightarrow K^*$ transitions are  suppressed substantially by the
approximately 'thermal' pion spectrum \cite{Fuchs}.

Furthermore, we have pursued the idea of Refs.  \cite{Sorge,Bleich}
that the $K^\pm$ spectra could be hardened by string-string
interactions, which increase the effective string tension $\sigma$ and
thus the probability to produce mesons at high $m_T$
\cite{Ko_AMPT,Bleich}.  In order to estimate the largest possible
effect of string-string interactions we have assumed that for two
overlapping strings the string tension $\sigma$ is increased by a
factor of two, for three overlapping strings by a factor of three etc.
Here the overlap of strings is defined geometrically assuming a
transverse string radius $R_s$, which according to the studies in Ref.
\cite{Geiss99} should be $R_s \leq$ 0.25 fm.  Based on these
assumptions (and $R_s$=0.25 fm), we find only a small increase of the
inverse slope parameters at AGS energies, where the string densities
are low.  At 160 $A\cdot$GeV the model gives a significant hardening of
the spectra by about 15\%, which, however, is still significantly less
than the effect observed in the data.
\begin{figure}[t]
\begin{center}
\begin{minipage}[l]{13.cm}
\epsfig{file=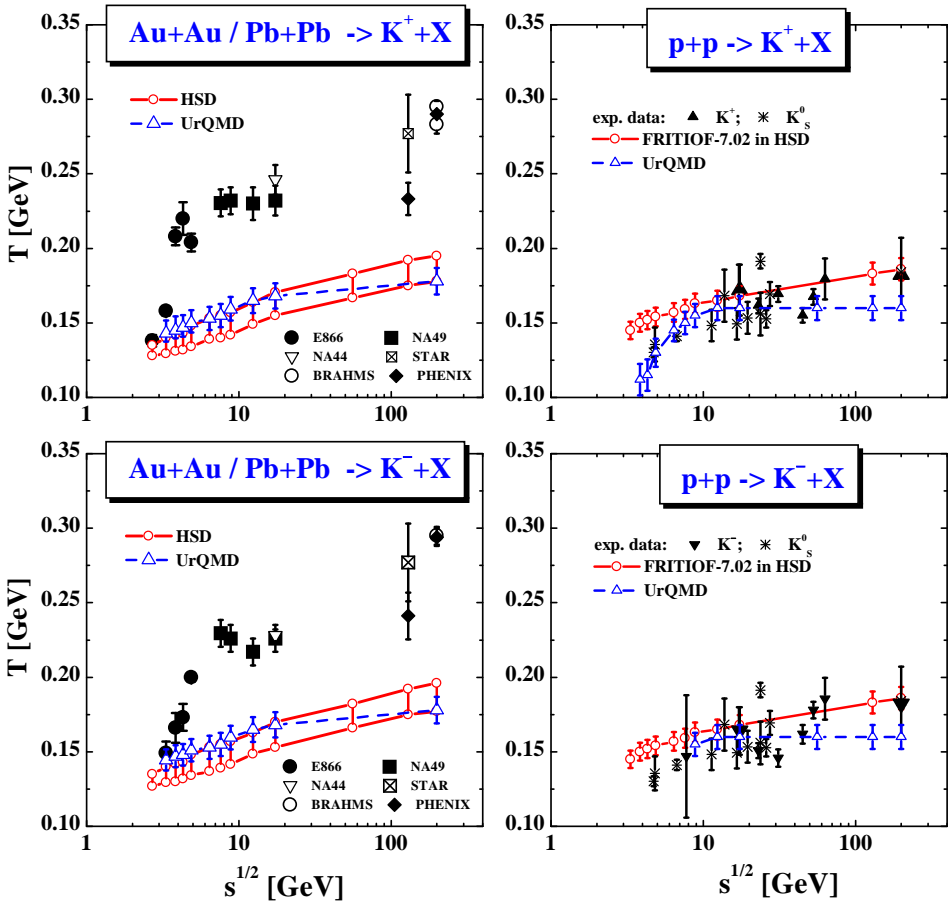,scale=0.65}
\end{minipage}
\begin{minipage}[l]{5.cm}
\caption{Comparison of the inverse slope parameters $T$ for $K^+$ and
$K^-$ mesons from central Au+Au (Pb+Pb) collisions (l.h.s.) and $pp$
reactions (r.h.s.) as a function of the invariant energy $\sqrt{s}$ from
HSD (upper and lower solid lines) and UrQMD (open triangles) with
data from Refs. \protect\cite{E866E917,NA49_T,NA44,STAR,BRAHMS,PHENIX}
for $AA$ and \protect\cite{NA49_CCSi,Gazdz_pp,STAR} for $pp$ collisions.
The upper and lower solid lines result from different limits of
the HSD calculations as discussed in the text.
The figure is taken from Ref. \protect\cite{Bratnew}.}
 \label{Fig_T}
\end{minipage}
\end{center}
\end{figure}

Our findings are summarized in Fig. \ref{Fig_T}, where the dependence of
the inverse slope parameter $T$ (see Eq.~(\ref{slope})) on $\sqrt{s}$
is shown and compared to the experimental data \cite{NA49_T,NA49_CCSi} for
central Au+Au (Pb+Pb) collisions (l.h.s.) and $pp$ reactions (r.h.s.).
The upper and lower solid lines (with open circles) on the l.h.s. in
Fig. \ref{Fig_T} correspond to results from HSD calculations, where the
upper and lower limits are due to fitting the slope $T$ itself, an
uncertainty in the repulsive $K^\pm$-pion potential or the possible
effect of string overlaps.
The slope parameters from $pp$ collisions (r.h.s. in Fig.  \ref{Fig_T})
are seen to increase smoothly with energy both in the experiment (full
squares) and in the HSD calculations (full lines with open circles).
The UrQMD results for $pp$ collisions are shown as open triangles
connected by the solid line and systematically lower than the slopes
from HSD at all energies.

We mention that the RQMD model \cite{Sorge} gives higher inverse slope
parameters for kaons at AGS and SPS energies than HSD and UrQMD, which
essentially might be traced back to the implementation of effective
resonances with masses above 2 GeV as well as 'color ropes' that decay
isotropically in their rest frame \cite{Hecke}.  A
more detailed discussion of this issue is presented in Ref.
\cite{Bratnew}.

\section{Thermodynamics in the $T-\mu_B$ plane}

This still leaves us with the question of the origin of the rapid
increase of the $K^\pm$ slopes with invariant energy for central Au+Au
collisions at AGS energies and the constant slope at SPS energies (the
'step'), which is  missed in both transport approaches. We recall that
higher transverse particle momenta either arise from repulsive self
energies -- in mean-field dynamics -- or from collisions, which reduce
longitudinal momenta in favor of transverse momenta \cite{HORST,CaMo}.
As shown above in Fig. \ref{Fig_T} conventional hadron self-energy
effects and hadronic binary collisions are insufficient to describe the
dramatic increase of the $K^\pm$ slopes as a function of $\sqrt{s}$.
This indicates additional mechanisms for the generation of the pressure
that is observed experimentally.

\begin{figure}[t]
\begin{center}
\begin{minipage}[l]{11.5cm}
\epsfig{file=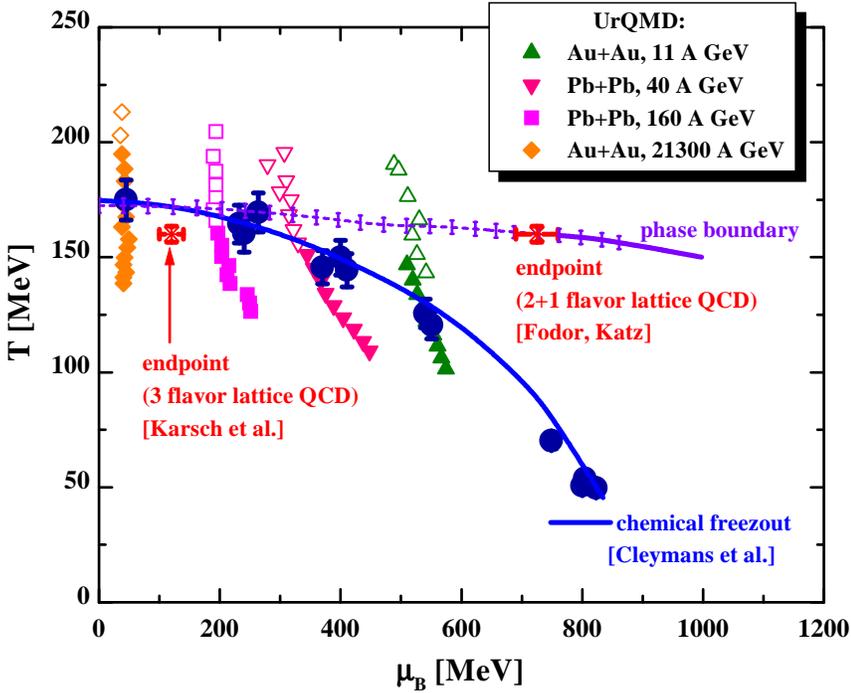,scale=0.6}
\end{minipage}
\begin{minipage}[l]{6.5cm}
\caption{ The solid line characterizes the universal chemical
freeze-out line from Cleymans et al. \protect\cite{Cleymans} whereas
the full dots (with errorbars) denote the 'experimental' chemical
freeze-out parameters from Ref. \protect\cite{Cleymans}. The various
symbols stand for temperatures $T$ and chemical potentials $\mu_B$
extracted from UrQMD transport calculations in central Au+Au (Pb+Pb)
collisions at 21.3 A$\cdot$TeV, 160, 40 and 11 A$\cdot$GeV
\protect\cite{Bravina} (see text). The stars indicate the tri-critical
endpoints from lattice QCD calculations by Karsch et al.
\protect\cite{Karsch2} (left point) and Fodor and Katz
\protect\cite{Fodor} (right point). The horizontal line with errorbars
is the phase boundary from \protect\cite{Fodor}.}
\label{Fig_QCD}
\end{minipage}
\end{center}
\end{figure}

Here we propose that additional pre-hadronic/partonic degrees of freedom might be
responsible for this effect already at $\sim$ 5 $A\cdot$GeV. Our
arguments are based on a comparison of the thermodynamic
parameters $T$ and $\mu_B$ extracted from the transport models in the
central overlap regime of Au+Au collisions \cite{Bravina} with the
experimental systematics on chemical freeze-out configurations
\cite{Cleymans} in the $T,\mu_B$ plane. The solid line in Fig.
\ref{Fig_QCD} characterizes the universal chemical freeze-out line from
Cleymans et al. \cite{Cleymans} whereas the full dots with errorbars
denote the 'experimental' chemical freeze-out parameters - determined
from the fits to the experimental yields - taken from Ref.
\cite{Cleymans}. The various symbols (in vertical sequence) stand for
temperatures $T$ and chemical potentials $\mu_B$ extracted from UrQMD
transport calculations in central Au+Au (Pb+Pb) collisions at 21.3
A$\cdot$TeV, 160, 40 and 11 A$\cdot$GeV \cite{Bravina} as a function of
the reaction time (from top to bottom).  The open symbols denote
nonequilibrium configurations and correspond to $T$ parameters
extracted from the transverse momentum distributions, whereas the full
symbols denote configurations in approximate pressure equilibrium in
longitudinal and transverse direction.

During the nonequilibrium phase (open symbols) the transport
calculations show much higher temperatures (or energy densities) than
the 'experimental' chemical freeze-out configurations at all bombarding
energies ($\geq$ 11 A$\cdot$GeV).  These numbers are also higher than
the tri-critical endpoints extracted from lattice QCD calculations by
Karsch et al.  \cite{Karsch2} and Fodor and Katz \cite{Fodor} (stars
with horizontal error bars). Though the QCD lattice calculations differ
substantially in the value of $\mu_B$ for the critical endpoint, the
critical temperature $T_c$ is in the range of 160 MeV in both
calculations, while the energy density is in the order of 1 GeV/fm$^3$
or even below. Nevertheless, this diagram shows that at RHIC energies
one encounters more likely a cross-over between the different phases
when stepping down in temperature during the expansion phase of the
'hot fireball'. This situation changes at lower SPS or AGS (as well as  new GSI
SIS-300) energies, where for sufficiently large chemical
potentials $\mu_B$ the cross over should change to a first order
transition \cite{Shuryak}, i.e., beyond the tri-critical point in the
($T,\mu_B$) plane.  Nevertheless, Fig. \ref{Fig_QCD} demonstrates that
the transport calculations show temperatures (energy densities) well
above the phase boundary (horizontal line with errorbars) in the very
early phase of the collisions, where hadronic interactions practically
yield no pressure, but pre-hadronic degrees of freedom should do. This
argument is in line with the studies on elliptic flow at RHIC
energies, that is underestimated by ~30\% at midrapidity in the HSD
approach for all centralities \cite{Brat03}. Only strong early
stage pre-hadronic interactions might
cure this problem.

\section{Conclusions}

Summarizing this contribution, we point out that baryon stopping
\cite{Weber_stop02} and hadron production in central Au+Au (or Pb+Pb)
collisions is quite well described in the independent transport
approaches HSD and UrQMD. Also the 'longitudinal' rapidity
distributions of protons, pions, kaons, antikaons and hyperons are
similar in both models and in  reasonable agreement with available
data. The exception are the pion rapidity spectra at the highest AGS
energy and lower SPS energies, which are overestimated by both models
\cite{Weber02}.  As a consequence the HSD and UrQMD transport
approaches underestimate the experimental maximum of the $K^+/\pi^+$
ratio ('horn') at $\sim$ 20 to 30 A$\cdot$GeV.
However, we point out that the maxima in the $K^+/\pi^+$ and
($\Lambda+\Sigma^0)/\pi$ ratios dominantly reflect a change from baryon
to meson dominated dynamics with increasing bombarding energy.

We have found that the inverse slope parameters $T$ for $K^\pm$ mesons
from the HSD and UrQMD transport models are practically independent of
system size from $pp$ up to central Pb+Pb collisions and show only a
slight increase with collision energy, but no 'step' in the $K^\pm$
transverse momentum slopes. The rapid increase of the inverse slope
parameters of kaons for collisions of heavy nuclei (Au+Au) found
experimentally in the AGS energy range, however, is not reproduced by
both models (see Fig.~\ref{Fig_T}).  Since the pion transverse mass
spectra -- which are hardly effected by collective flow  -- are
described sufficiently well at all bombarding energies
\cite{Bratnew}, the failure has to be attributed to a lack of pressure.
We have argued - based on lattice QCD calculations at finite
temperature and baryon chemical potential $\mu_B$ \cite{Fodor,Karsch2}
as well as the experimental systematics in the chemical freeze-out
parameters \cite{Cleymans} - that this additional pressure should be
generated in the early phase of the collision, where the 'transverse'
energy densities in the transport approaches are higher than the
critical energy densities for a phase transition (or cross-over) to the QGP. The
interesting finding of our analysis is, that pre-hadronic degrees of
freedom might already play a substantial role in central Au+Au collisions
at AGS energies above $\sim$~5~$A\cdot$GeV.

We recall that  the systematic studies of in-plane and elliptic
proton flow in Au+Au collisions at AGS energies \cite{Dani,Sahu,Toneev}
also indicate a 'softening' of the nuclear equation of state (EoS) at
4-6 A$\cdot$GeV, which further supports our present findings.


\begin{thebibliography}{99}
\itemsep -2pt
\bibitem{Rafelski}
    J.~Rafelski and B.~M\"uller, {\em Phys. Rev. Lett.} {\bf 48} (1982) 1066.
\bibitem{bjorken}
    J.D. Bjorken, {\em Phys. Rev.} D {\bf 27} {1983} 140.
\bibitem{QM01}
    {\it Quark Matter 2003}, {\em Nucl. Phys.} A {\bf 715} (2003) 1.
\bibitem{Karsch}
    F. Karsch {\it et al.}, {\em Nucl. Phys.} B {\bf 502} (2001) 321.
\bibitem{HORST}
    H. St\"ocker and W. Greiner, {\em Phys. Rep.} {\bf 137} (1986) 277.
\bibitem{exita}
    W. Cassing, E.L. Bratkovskaya, and S. Juchem,
    {\em Nucl. Phys.} A {\bf 674} (2000) 249.
\bibitem{Fodor}
    Z. Fodor and S. D. Katz,  {\em JHEP} {\bf 0203} (2002) 014.
% preprint hep-lat/0208078.
\bibitem{NA49_T}
    V. Friese {\it et al.}, NA49 Collaboration, preprint nucl-ex/0305017.
\bibitem{Goren}
    M. I. Gorenstein, M. Ga\'zdzicki, and K. Bugaev,
    {\em Phys. Lett.} B {\bf 567} (2003) 175.
\bibitem{SMES}
    M. Ga\'zdzicki and M. I. Gorenstein,
    {\em Acta Phys. Polon.} B {\bf 30} (1999) 2705.
\bibitem{Marek}
    M. Ga\'zdzicki, this volume.
\bibitem{Hove}
    L. Van Hove, {\em Phys. Lett.} B {\bf 118} (1982) 138.
\bibitem{UrQMD1}
    S.A.~Bass {\it et al.},
    {\em Prog. Part. Nucl. Phys.} {\bf 42} (1998) 255.
\bibitem{UrQMD2}
    M.~Bleicher {\it et al.},  {\em J. Phys.} G {\bf 25} (1999) 1859.
\bibitem{Geiss}
    J. Geiss, W. Cassing, and C. Greiner,
    {\em Nucl. Phys.} A {\bf 644} (1998) 107.
\bibitem{Cass99}
    W. Cassing and E. L. Bratkovskaya,
    {\em Phys. Rep.} {\bf 308} (1999) 65.
\bibitem{Weber02}
    H. Weber, E.L. Bratkovskaya, W. Cassing, and H. St\"ocker,
    {\em Phys. Rev.} C {\bf 67} (2003) 014904.
\bibitem{PDG}
    K.~Hagiwara {\it et al.}, (Review of Particle Properties),
    {\em Phys. Rev.} D {\bf 66} (2002) 010001.
\bibitem{LUND}
    B. Andersson {\it et al.}, {\em Z. Phys. C} {\bf 57} (1993) 485.
\bibitem{Ko_AMPT}
     Z. W. Lin {\it et al.}, {\em Nucl. Phys.} A {\bf 698} (2002) 375.
\bibitem{E866E917}
    L. Ahle {\it et al.}, E866 and E917 Collaboration,
    {\em Phys. Lett.} B {\bf 476} (2000) 1;
    {\it ibid.} {\bf 490} (2000) 53.
\bibitem{E895} % new pion spectra
    J. L. Klay {\it et al.}, E895 Collaboration,
    preprint nucl-ex/0306033.
\bibitem{E891Lam}
      S. Ahmad {\it et al.}, E891 Collaboration,
    {\em Phys. Lett.} B {\bf 382} (1996) 35;
      C. Pinkenburg {\it et al.}, E866 Collaboration,
        {\em Nucl. Phys.} A {\bf 698} (2002) 495c.
\bibitem{NA49_new}
     S. V. Afanasiev {\it et al.}, NA49 Collaboration,
     {\em Phys. Rev.} C {\bf 66} (2002) 054902.
\bibitem{NA49_Lam} % (Lambdas)
    A.~Mischke {\it et al.}, NA49 Collaboration,
    {\em J. Phys.} G. {\bf 28} (2002) 1761;
     {\em Nucl. Phys.} A {\bf 715} (2993) 453.
\bibitem{Antiori} % (Lambdas)
    F. Antinori {\it et al.}, WA97 Collaboration,
    {\em Nucl. Phys.} A {\bf 661} (1999) 130c.
\bibitem{BRAHMS}
    D. Ouerdane {\it et al.}, BRAHMS Collaboration,
    {\em Nucl. Phys.} A {\bf 715} (2003) 478;
    J. H. Lee {\it et al.}, to be published in the Proceedings of
      {\em 'Strange Quark Matter-03'}.
\bibitem{PHENIX}
    S. S. Adler {\it et al.}, PHENIX Collaboration,
    preprint nucl-ex/0307010; preprint nucl-ex/0307022.
\bibitem{STAR}
    C. Adler {\it et al.}, STAR Collaboration, preprint nucl-ex/0206008;
    O. Barannikova {\it et al.}, {\em Nucl. Phys.} A {\bf 715} (2003) 458;
    K. Filimonov {\it et al.}, preprint hep-ex/0306056.
\bibitem{Bratnew}
    E. L. Bratkovskaya {\it et al.}, to be published.
\bibitem{Brat03}
    E. L. Bratkovskaya, W. Cassing and H. St\"ocker,
    {\em Phys. Rev.} C {\bf 67} (2003) 054905.
\bibitem{ERICE2}
    W. Cassing, K. Gallmeister, E. L. Bratkovskaya,
    C. Greiner, and H. St\"ocker, this volume.
\bibitem{Soff03}
    S. Soff {\it et al.}, {\em Phys. Lett.} B {\bf 551} (2003) 115.
\bibitem{Xipi}
    K. Redlich, J. Cleymans, H. Oeschler, and A. Tounsi,
    {\em Acta Phys. Polonica} B {\bf 33} (2002)1609.
\bibitem{Brat03PRL}
    E. L. Bratkovskaya, S. Soff, H. St\"ocker, M. van Leeuwen, and
    W. Cassing, preprint nucl-th/0307098, {\em Phys. Rev. Lett.}, in press.
\bibitem{KaoS}
    A. F\"orster {\it et al.}, KaoS Collaboration,
    {\em J. Phys.} G {\bf 28} (2002) 2011.
\bibitem{Fuchs}
    B. V. Martemyanov {\it et al.}, nucl-th/0212064.
\bibitem{Sorge}
    H. Sorge, {\em Phys. Rev.} C {\bf 52} (1995) 3291.
\bibitem{Bleich} % (string enhancement)
      S. Soff {\it et al.}, {\em Phys. Lett.} B {\bf 471} (1999) 89.
\bibitem{Geiss99}
    J. Geiss {\it et al.}, {\em Phys. Lett.} B {\bf 447} (1999) 31.
\bibitem{NA49_CCSi}
    I. Kraus {\it et al.}, NA49 Collaboration,
    preprint nucl-ex/0306022.
\bibitem{NA44} % T at 160 GeV/A
    I.G. Bearden {\it et al.}, NA44 Collaboration,
    preprint nucl-ex/0202019.
\bibitem{Gazdz_pp}  % compilation on T-slope from pp
    M. Kliemant, B. Lungwitz, and M. Ga\'zdzicki,
    preprint hep-ex/0308002.
\bibitem{Hecke} % RQMD
    H. van Hecke {\it et al.}, {\em Phys. Rev. Lett.} {\bf 81} (1998) 5764.
\bibitem{CaMo}
    W. Cassing and U. Mosel, {\em Prog. Part. Nucl. Phys.}
    {\bf 25} (1990) 235.
\bibitem{Bravina}
    L. V. Bravina {\it et al.}, {\em Phys. Rev.} C {\bf 60} (1999) 024904;
    {\em Nucl. Phys.} A {\bf 698} (2002) 383.
\bibitem{Cleymans}
    J. Cleymans and K. Redlich, {\em Phys. Rev.} C {\bf 60} (1999) 054908.
\bibitem{Karsch2}
    F. Karsch {\it et al.}, preprint hep-lat/0309116.
\bibitem{Shuryak}
    E. Shuryak, this volume.
\bibitem{Weber_stop02}
      H. Weber, E.L. Bratkovskaya and H. St\"ocker,
    {\em Phys. Lett.} B {\bf 545} (2002) 285.
\bibitem{Dani}
    P. Danielewicz {\it et al.},
    {\em Phys. Rev. Lett.} {\bf 81} (1998) 2438.
\bibitem{Sahu}
    P. K. Sahu and W. Cassing, {\em Nucl. Phys.} A {\bf 712} (2002) 357.
\bibitem{Toneev}
    E.G. Nikonov, A.A. Shanenko and V.D. Toneev,
    {\em Heavy Ion Phys.} {\bf 8} (1998) 89.

\end{thebibliography}
\end{document}